# Hall sensors batch-fabricated on all-CVD h-BN/graphene/h-BN heterostructures

André Dankert[†,*], Bogdan Karpiak[*], Saroj P. Dash[‡]

*Department of Microtechnology and Nanoscience, Chalmers University of Technology,*

*SE-41296, Göteborg, Sweden.*

The two-dimensional (2D) material graphene is highly promising for Hall sensors due to its potential of having high charge carrier mobility and low carrier concentration at room temperature. Here, we report the scalable batch-fabrication of magnetic Hall sensors on graphene encapsulated in hexagonal boron nitride (h-BN) using commercially available large area CVD grown materials. The all-CVD grown h-BN/graphene/h-BN van der Waals heterostructures were prepared by layer transfer technique and Hall sensors were batch-fabricated with 1D edge metal contacts. The current-related Hall sensitivities up to 97 V/AT are measured at room temperature. The Hall sensors showed robust performance over the wafer scale with stable characteristics over six months in ambient environment. This work opens avenues for further development of growth and fabrication technologies of all-CVD 2D material heterostructures and allows further improvements in Hall sensor performance for practical applications.

**Introduction**

Magnetic field sensors today represent a significant growing market, estimated to reach USD 4.16 billion in value terms by the year 2022[1]. The areas of application cover many fields, such as automotive, consumer electronics, healthcare and defense industry, where magnetic field sensors are used for position detection, current monitoring and angular sensing. Many different magnetic sensors based on a variety of effects have been realized for different applications[2]. Hall effect-based sensors constitute a significant part of the industry, with an estimated market share to be over 55% in 2014[3]. They are used for magnetic field detection in the field range from $10^{-7}$ T to $10^2$ T in a temperature range from -40℃ up to 150℃. Today, the most ubiquitous sensors utilize an active region made of Si due to the low fabrication cost, highly developed processing technology, good integration into signal processing circuits and reasonable performance properties (current-normalized sensitivity $S_I$~100 V/AT)[4–7]. In comparison, Hall sensors based on III-V compound semiconductors provide better performances[8–11], but are expensive and more difficult to integrate in circuits.

Graphene is highly interesting material to be used as active region of magnetic Hall sensors, owing to its 2D nature, low carrier concentration $n_{2D}$ and high carrier mobility $\mu$. The previous reports on graphene Hall sensors demonstrated current-related sensitivities ($S_I \propto 1/n_{2D}$) up to 1200 V/AT on large area unprotected CVD graphene on $SiO_2$ substrate[12] and 1020 V/AT on epitaxial graphene on SiC substrates[13]. However, single layer graphene devices are prone to contaminations from environment and encapsulated structures are needed for reliable and durable performance for practical applications[14]. Recent studies on graphene encapsulation by $Al_2O_3$ grown by atomic layer deposition showed good results with low doping levels[15]. However, the encapsulated graphene with 2D insulating h-BN flakes provide superior interface, containing low amount of dangling bonds and charge traps and retaining the high electronic properties of graphene for high Hall sensor performance[16]. Moreover, the smoothness of h-BN allows using it as a high-quality substrate in addition to top encapsulation. Thus, current-related sensitivities $S_I$~5700 V/AT were obtained on stacks of all-exfoliated h-BN/graphene/h-BN[17], $S_I$~2270 V/AT on single crystal CVD graphene patches encapsulated between exfoliated h-BN[18], and $S_I$~1986 V/AT with batch-fabricated CVD graphene on in-situ grown CVD h-BN substrate without top encapsulation[19]. In addition to increased sensitivities, high linearity[12,20] and low noise of the devices[20–22] combined with transparency and flexibility[18,23] generates a high interest in graphene for the use in magnetic Hall sensors. However, for practical utilization in ambient environment over longer period, there is a necessity to investigate graphene Hall sensors fabricated by a scalable process using all-CVD grown 2D material heterostructures with full encapsulation of graphene active region.

Here we report a scalable graphene Hall sensor fabrication process by using commercially available large area all-CVD grown 2D materials van der Waals heterostructures. The devices used here involve an active region consisting of CVD graphene, which is encapsulated with CVD grown multilayer h-BN. The 1D edge metal contacts are utilized to connect the graphene active region to the electronic circuit[24]. The Hall sensors exhibit stable performance over the wafer scale in ambient environmental conditions over a longer period of time. Although further improvements are required in the fabrication process, such large-scale batch fabrication of h-BN encapsulated graphene Hall elements can bring the technology closer to practical applications.

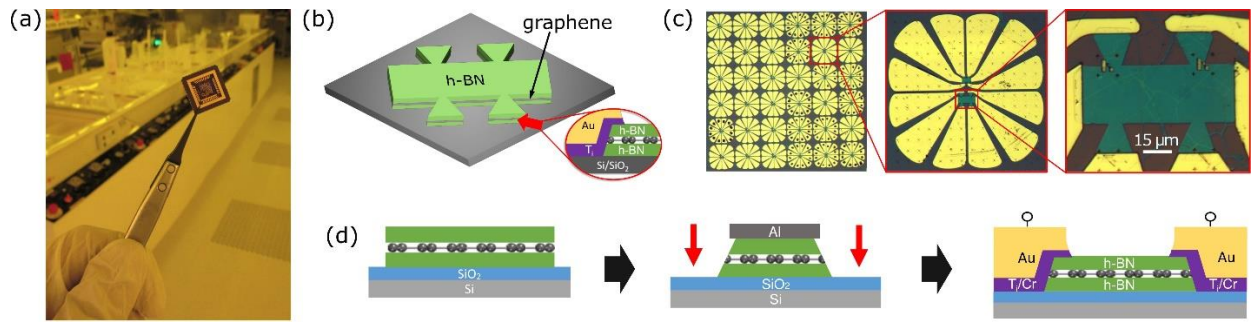

*Figure 1:* *Fabrication of magnetic Hall elements based on large area CVD graphene encapsulated in CVD h-BN on Si/SiO$_2$ substrate. (a) Picture of the chip carrier with batch-fabricated graphene Hall elements. (b) The schematic representation of the fabricated Hall sensor devices with h-BN/graphene/h-BN heterostructure and 1D edge contacts. (c) Optical microscope picture of the batch-fabricated chip and individual graphene Hall element. (d) Schematic of the fabrication process steps involving the preparation of h-BN/graphene/h-BN heterostructures by layer transfer method, followed by patterning and formation of 1D edge metal contacts.*

**Results**

The Hall sensors were prepared by using a scalable fabrication process. An optical picture of a chip-size batch-fabricated graphene Hall elements is shown in Fig. 1a. Figure 1b shows the schematic representation of the Hall bar devices with 1D edge contacts. Figure 1c presents an optical microscope picture of the chip and an individual graphene Hall element, which were fabricated through the micro-fabrication process as schematically shown in Fig. 1d. Each Hall element consists of CVD graphene (from Graphenea[25]) sandwiched between multilayer CVD h-BN (from Graphene Supermarket[26]). The 2D material heterostructure was prepared on a Si/SiO$_2$ wafer by large area PMMA-supported wet-transfer technique and Ar/H$_2$ annealing for each layer (service of Graphenea). Next, they were patterned to Hall bar structures by using an Al (20 nm) hard mask for etching with CHF$_3$ and O$_2$ gas. The Al hard mask was removed by wet chemical etching in Shipley Microposit MF-319. The 1D edge contacts were fabricated by means of photo-lithography and electron beam evaporation of metals followed by liftoff in acetone. More specifically, we investigated 1D edge metal contacts of Cr (5 nm)/Au (95 nm) and Ti (20 nm)/Au (60 nm) to 2D graphene channels.

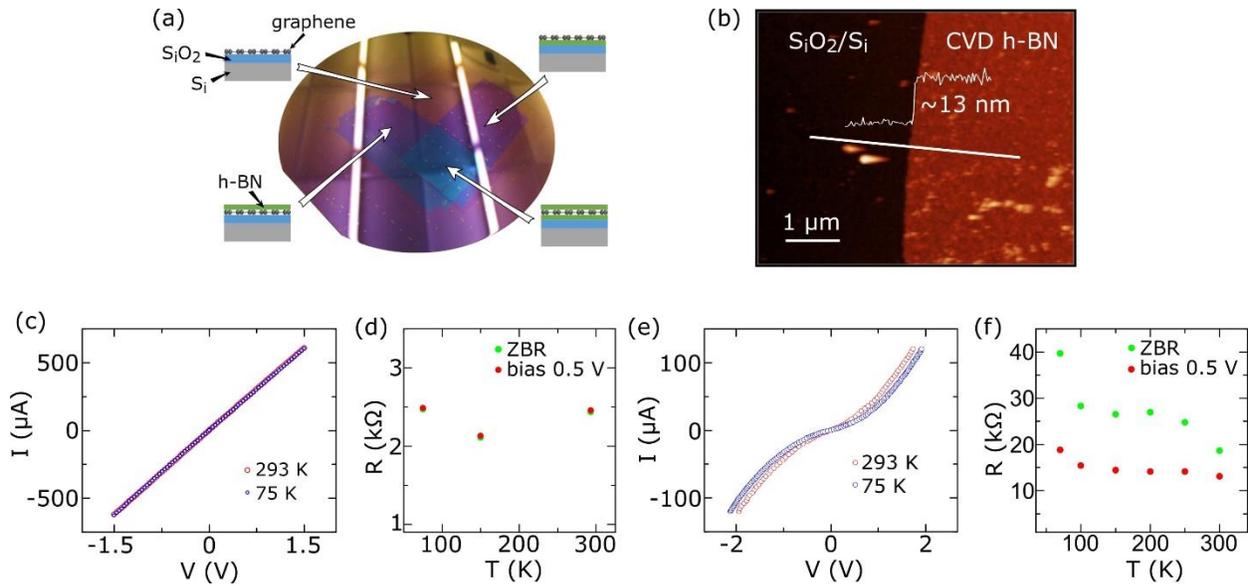

*Figure 2:* Characterization of the 2D heterostructures and 1D contacts. (a) The prepared h-BN/graphene/h-BN heterostructures using all-CVD grown 2D materials on a 4-inch SiO$_2$/Si wafer by layer transfer method. Different regions containing 2D layers and heterostructures are indicated by schematics. (b) AFM image and thickness profile of CVD h-BN on SiO$_2$/Si wafer. (c) Two-terminal IV characteristic of the device with Cr/Au edge contacts to graphene at 293 K and 75 K. (d) Cr/Au edge contact resistance at 0.5 V bias (red) and zero bias voltage (green) as a function of temperature. (e) 2-terminal IV characteristic of the device with Ti/Au edge contacts to graphene at 293 K (red) and 75 K (blue). (f) Ti/Au edge contact resistance at 0.5 V bias (red) and zero bias voltage (green) as a function of temperature.

Figure 2a shows the prepared all-CVD h-BN/graphene/h-BN heterostructures on a 4-inch SiO$_2$/Si wafer. The Raman spectrum of monolayer CVD graphene[27] on SiO$_2$/Si substrate show the G and 2D peaks at 1597 cm$^{-1}$ and 2652 cm$^{-1}$ respectively[28] with a small D peak (see Supplementary Fig. S1a). The grain size of the CVD graphene is mostly between 1-5 µm range. The Raman characterization of the h-BN film revealed a peak at 1357 cm$^{-1}$ at selective places[29] (see Supplementary Figure S1b). The thickness of the CVD h-BN used in the heterostructures was measured by AFM (~10-13 nm, as shown in Fig. 2b). The rms roughness of the h-BN films were found to be 1 - 2 nm on Cu foil and SiO$_2$ substrate. Although we could get rid of organic contamination introduced on h-BN from the transfer and device fabrication process by annealing in Ar/H$_2$, the roughness remains at similar values (see Supplementary Fig. S2). Earlier, Kim et al reported the multilayer CVD h-BN films to be polycrystalline in nature and indicated sp$^2$ bond coordination of B and N atoms[30].

The electrical characteristics of 1D edge contacts with Cr/Au and Ti/Au metals to graphene are shown in Fig. 2c-f. As observed from current-voltage (IV) characteristics, the Cr/Au contacts provide a low-resistive linear behavior, while Ti/Au contacts show a high-resistive non-linear tunneling behavior. The weak temperature dependence of the resistance for the Cr/Au contacts at zero bias and at 0.5 V bias (Fig. 2d) indicates high quality interfaces[31]. However, the high resistance and tunneling behavior of the Ti/Au edge contact to graphene (Fig. 2e and Fig. 2f) could be due to interfacial species, such as oxidation at the interfaces. The encapsulated CVD graphene is found to be hole-doped with sheet resistances $R_S$ between 520 - 870 Ω/□.

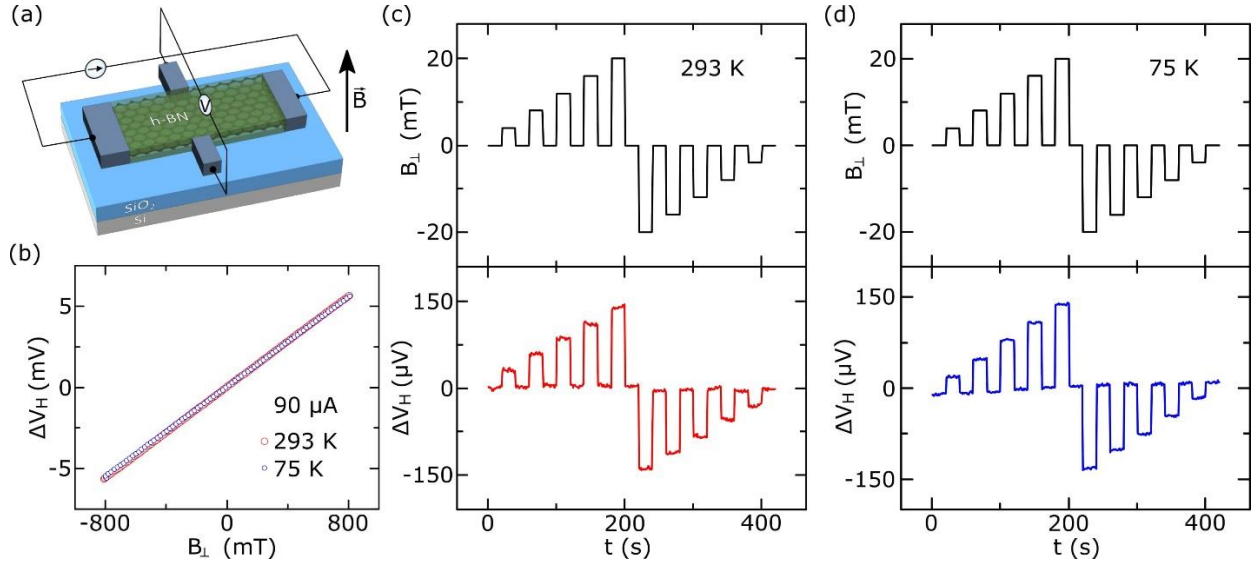

*Figure 3:* Characterization of the graphene Hall elements with Cr/Au 1D edge contacts. (a) The schematic of Hall measurements. (b) Hall voltage as a function of perpendicular magnetic field at 293 K and 75 K. (c) Hall voltage as a function of time at different applied magnetic fields at 293 K and (d) 75 K. Measurements are shown for a bias current of I=90 μA and the background offset voltage is subtracted from the measured data.

First, we present the characterization of the h-BN/graphene/h-BN Hall sensors with Cr/Au edge contacts. The samples were measured at ambient conditions, i.e. at room temperature of 293 K, pressure of 750 Torr, relative humidity ~ 70%. Except for the cases where measurements were done at room temperature, pressure was $10^{-2}$ Torr. The Hall voltage $V_H$ is measured at a constant applied current, while sweeping a perpendicular magnetic field (Fig. 3a). The Lorentz force acting on the moving charges in graphene resulted in a voltage difference in the transverse direction (Hall voltage $V_H$). The detected Hall voltage, measured as a function of applied perpendicular magnetic field $B_\perp$ at bias current I=90 μA, is shown in Fig 3b. A very weak temperature-dependent change is observed within the range of 75-293 K. A background voltage offset $V_{offset}$, due to a misalignment between Hall probe contacts, has been subtracted. Fitting the Hall response with $V_H = \frac{IB_\perp}{en_{2D}} + V_{\text{offset}}$, where e is the electron's elementary charge and $n_{2D}$ is the sheet charge carrier concentration[4,11], we can extract the Hall mobilities $\mu = 1/(e|n_{2D}|R_S)$ around 1200 cm²/Vs$^{-1}$ with carrier concentrations around $8 \times 10^{12}$ cm$^{-2}$ at zero gate voltage. The high charge doping is most likely due to the wet transfer process that might have trapped impurities at the h-BN/ graphene interfaces. The linearity error is found to be around 0.25% and independent of temperature. The Hall sensors also showed stable performance and good response to magnetic field changes in time (Fig. 3c and Fig. 3d).

In order to investigate the effect of the 1D edge contact material on Hall sensor performance, we also carried out measurements on devices with Ti/Au contacts. The output Hall voltage as a function of applied magnetic field measured at 293 K and 75 K with a bias current of I=90 μA is shown in Fig 4a. Despite an order of magnitude higher contact resistance in Ti/Au devices, we observed a similar Hall sensitivity as in Cr/Au devices. The linearity errors in Hall measurements were found to be at an average level of ~2 % at 293 K down to ~0.4 % at 75 K, which is 2 to 8 times higher than for the Cr/Au device. Figures 4b and Fig. 4c show the Hall voltage responses measured as a function of time at different perpendicular magnetic fields at 293 K and 75 K, respectively. This shows that Cr/Au edge contacts perform better in terms of higher linearity of the Hall response and lower noise compared to Ti/Au. Noise spectral characterization of device with Cr/Au contacts (see Supplementary Fig. S3) revealed minimum magnetic resolution of 0.4 mT/Hz$^{0.5}$.

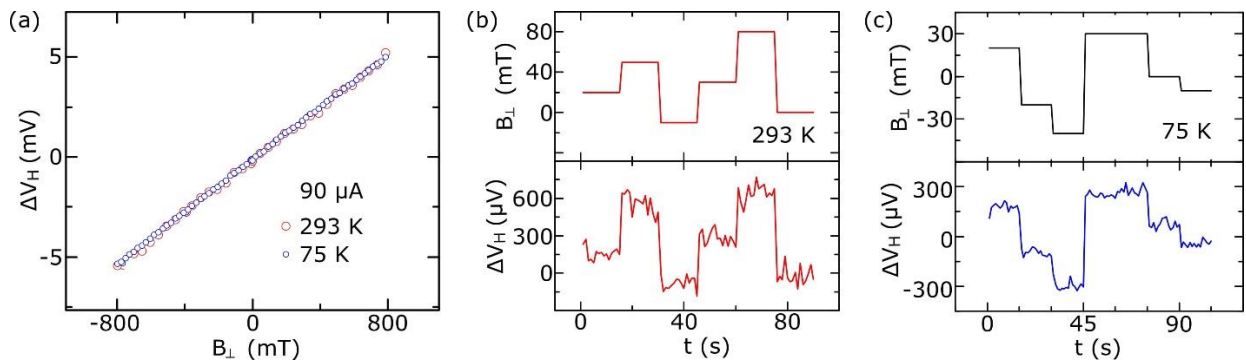

*Figure 4:* *Characterization of the graphene Hall elements with Ti/Au 1D edge contacts. (a) Output Hall voltage as a function of perpendicular magnetic field strength at 293 K and 75 K. (b) Output Hall voltage as a function of time at different applied magnetic fields at 293 K and (c) 75 K. Measurements are shown at a bias current of 90 μA and the background voltage offset is subtracted from the data.*

From the Hall measurements the current-related sensitivity $S_I = (\partial_B V_H)/I$ was extracted. Figure 5a shows the temperature dependence of $S_I$ for the Cr/Au and Ti/Au contacts, which are found to be thermally stable with the values of $S_I$~75 V/AT. The absence of temperature variation of $S_I$ ($S_I$~$1/n_{2D}$) is due to a stable carrier concentration ($n_{2D}$) in graphene. Figure 5b shows the current bias dependence with slight decrease of $S_I$ at higher bias, which could be due to heating-related effects[19]. The more prominent bias-induced decrease of $S_I$ with Ti/Au contacts compared to Cr/Au indicates more heating-related effects due to higher contact resistances. Comparing several different devices yields a narrow distribution of Hall sensitivity $S_I$ (Fig. 5c) ranging from 60 – 97 V/AT. Furthermore, variations in geometry and size of the graphene Hall elements did not affect the sensitivity, with the $S_I$ values ranging from 65 – 78 V/AT at room temperature (Fig. 5d). From the Hall measurement at constant voltage bias of 250 mV the voltage-related sensitivity for device with Cr/Au contacts was found to be 0.03 V/VT.

To compare the results of all-CVD h-BN/graphene/h-BN with differently prepared graphene Hall sensors, we have fabricated control samples of CVD graphene on $SiO_2$ substrate with and without exfoliated h-BN capping (see Supplementary Fig. S4). The CVD graphene devices on $SiO_2$ substrate without any encapsulation show $S_I$ in the range of 100 – 200 V/AT at room temperature. However, we observed degradation in sensitivity with time showing values below 65 V/AT after few weeks of fabrication. The Hall sensors with exfoliated h-BN encapsulation show the current-related sensitivity up to 363 V/AT at room temperature, which is 3 to 4 times higher than the all-CVD heterostructure devices. For the control CVD graphene samples, the Hall mobility and carrier concentrations are found to be around 150 $cm^2/Vs^{-1}$ and $2 \times 10^{12}$ $cm^{-2}$, respectively, which is lower in comparison to all-CVD heterostructures prepared by wet transfer process. The higher carrier concentrations due to trapped impurities at the h-BN/graphene interfaces are supposed to give rise to lower Hall sensitivity in our all-CVD h-BN/graphene/h-BN Hall sensors.

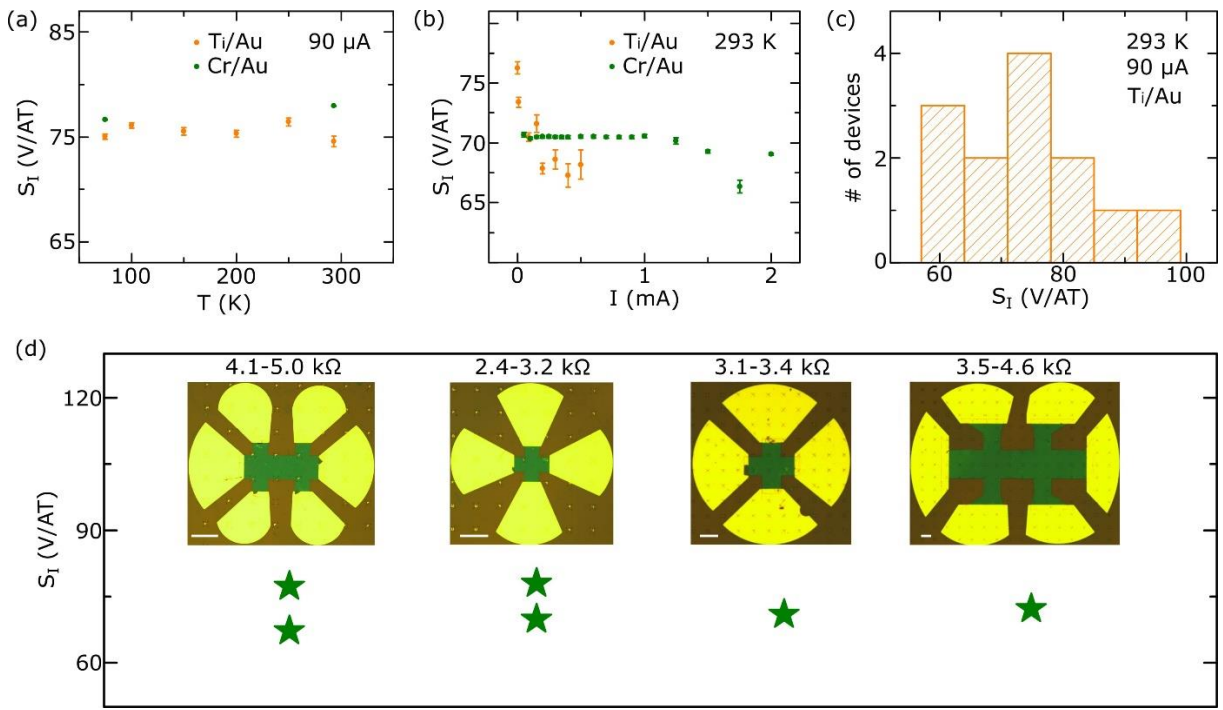

*Figure 5:* Current-related sensitivities ($S_I$) of graphene Hall elements. (a) Temperature dependence of $S_I$ for Cr/Au and Ti/Au contacts measured at 90 µA current bias. (b) Current bias dependence of $S_I$ for Cr/Au and Ti/Au contacts at room temperature. (c) Distribution of sensitivities $S_I$ measured in 13 Hall elements with Ti/Au contacts at 90 µA current bias and at room temperature. The devices were selected from different parts of the wafer. (d) Sensitivities $S_I$ for Hall sensors measured on devices with different size and geometry (Cr/Au contacts). Typical two-terminal resistances are shown with average resistivity value of $\sim 75 - 100$ $k\Omega \cdot \mu m$ (graphene sheet concentration $\sim 8.5 \cdot 10^{12}$ $cm^{-2}$). Scale bars in the inset images are 60 µm.

In order to verify the stability of our all-CVD h-BN/graphene/h-BN Hall sensors over time in an ambient environment, we have carried out measurements up to 190 days after fabrication. The Hall voltages measured after 1, 22, 49 and 190 days from device fabrication are shown in Fig 6a. The Hall sensors showed a good response to magnetic field changes even after 190 days (Fig. 6b) and a consistent Hall sensitivity $S_I$ of 53 – 78 V/AT at room temperature (Fig. 6c), without much degradation of the contacts and the graphene channel over time. However, unencapsulated graphene showed fast degradation of sensitivity with time after fabrication of devices. The atomically thin graphene is known to degrade quickly in ambient conditions not only due to electronic doping, but also chemical and mechanical damages due to exposure to different environmental conditions[32,33]. Although the high doping of graphene in our devices can contribute to less degradation of Hall sensitivity over time, the utilized h-BN encapsulation is imperative step forward towards protection of graphene for practical applications in ambient environment.

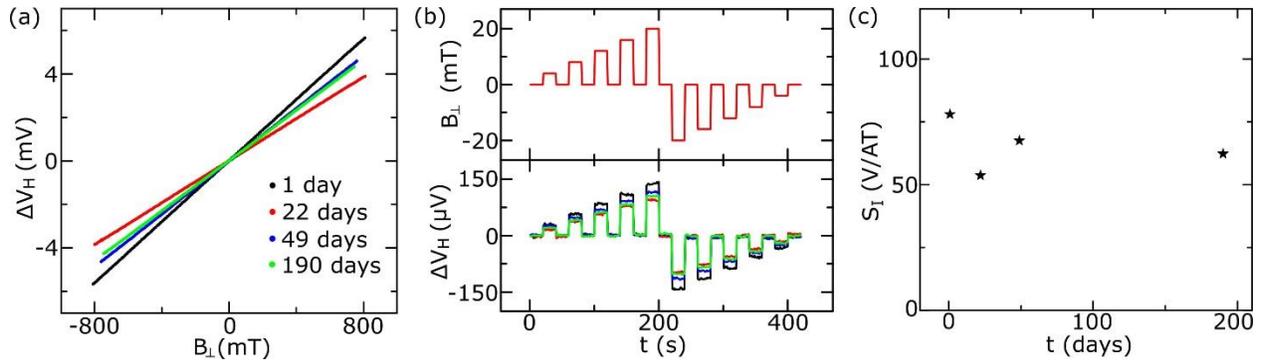

*Figure 6:* Stability of all-CVD h-BN/Graphene/h-BN Hall sensors over time in ambient environment. (a) Hall voltage as a function of magnetic field and (b) time at different applied magnetic fields at 293 K after keeping the device in ambient environment for 1 day (black), 22 days (red), 49 days (blue) and 190 days (green). (c) Current-related Hall sensitivity $S_I$ over time.

**Discussion**

Our observed current-related Hall sensitivities in large area all-CVD h-BN encapsulated graphene devices are comparable to those in magnetic Hall sensors based on Si[4–7]. However, sensitivities were at least one order of magnitude below exfoliated graphene/h-BN structures[17], unprotected CVD graphene devices on Si/SiO$_2$ substrate[12] and on CVD h-BN substrate[19]. Also, our control experiment with CVD graphene on SiO$_2$ substrate with/without exfoliated h-BN capping showed 2-4 times higher sensitivity. However, the unprotected graphene showed faster degradation of sensitivity with time. The higher sensitivity of devices with exfoliated h-BN encapsulation can be due to introduction of roughness, ripples and wrinkles in the CVD graphene in all-CVD h-BN/graphene/h-BN heterostructures. The CVD grown multilayer h-BN films are known to be polycrystalline in nature and the c-axis of the crystallites points to random directions[30], giving rise to defects and roughness in the heterostructures. The significant doping of graphene in the all-CVD heterostructure devices is another major reason behind the reduced Hall sensitivities and it could be due to contaminations introduced during wet-transfer process of large area graphene and h-BN layers. This could explain also the significantly lower minimum magnetic field resolution of Hall sensors based on all-CVD h-BN/graphene/h-BN heterostructures compared to previous reports on graphene Hall sensors[17,19–22]. These are currently the practical challenges for the development of graphene and 2D materials science and technology.

Such Hall sensors would greatly benefit from improved CVD growth methods of graphene and multi-layer h-BN, and large area layer transfer techniques for fabrication of heterostructures. Furthermore, development of methods for in-situ growth[34–36] of high-quality h-BN/graphene/h-BN van der Waals heterostructures on large areas would by-pass fabrication-related problems and allow for higher Hall sensitivities with lower graphene doping levels. Additionally, the contacts resistance in 1D geometry can also play an important role on device performance. Our reported reduction in contact resistance by switching from Ti to Cr can be further improved by optimizing the fabrication process, in particular the etching angle and 1D edge contact deposition to eliminate incorporated species at the interfaces[24].

**Conclusion**

In summary, we demonstrated h-BN/graphene/h-BN van der Waals heterostructure Hall sensors using all-CVD 2D materials available on the market. The batch-fabricated Hall sensors with Cr/Au 1D edge contacts showed reproducible contact properties with low resistances. Hall measurements and time-dependent response at different applied magnetic fields revealed current-related Hall sensitivities in the range of 60-97 V/AT at room temperature. Such encapsulated Hall sensors also showed durable operations in ambient environmental conditions over six months. This study demonstrates proof-of-concept batch fabrication of fully encapsulated all-CVD graphene Hall elements allowing for further development towards practical applications.

**Acknowledgements**

We acknowledge other financial support from EU Graphene Flagship (No. 604391), EU FlagEra project (No. 2015-06813), Swedish Research Council grants (No. 2012-04604 and No. 2016-03658), Graphene center and AoA Nano program at Chalmers University of Technology. AD acknowledges financial support from the Chalmers Innovation office. BK acknowledges scholarship from EU Erasmus Mundus Nanoscience and Nanotechnology master programme. We acknowledge the service by Graphenea for the layer transfer of 2D materials and support from Chalmers Nanofabrication facility.

**Author contributions -** A.D. and B.K. carried out the device fabrication and measurements, and equally contributed. S.P.D. supervised the research. B.K. wrote the manuscript with input from all authors.

**Additional Information**

Authors declare no competing financial interests. Supplementary information accompanies this paper at https://www.nature.com/srep/.

## References

1. Magnetic Field Sensors Market by Type (Hall Effect, Magnetoresistive (AMR, GMR, TMR), MEMS-based, SQUID, Fluxgate)), Range (10 gauss), Application, End-User Industry - Global Forecasts to 2022. *Markets and Markets* (2016).

2. Díaz-Michelena, M. Small magnetic sensors for space applications. *Sensors* **9,** 2271–2288 (2009).

3. Magnetic Sensors Market Analysis By Technology ( Hall Effect Sensing , AMR , GMR ), By Application ( Automotive , Consumer Electronics , Industrial ) And Segment Forecasts To 2022. *Grand View Research* (2016).

4. Popovic, R. S. *Hall Effect Devices*. (Institute of Physics, 2003).

5. Boero, G., Demierre, M., Besse, P. A. & Popovic, R. S. Micro-Hall devices: Performance, technologies and applications. *Sensors Actuators A* **106,** 314–320 (2003).

6. Vervaeke, K., Simoen, E., Borghs, G. & Moshchalkov, V. V. Size dependence of microscopic Hall sensor detection limits. *Rev. Sci. Instrum.* **80,** 74701 (2009).

7. Kejik, P., Boero, G., Demierre, M. & Popovic, R. S. An integrated micro-Hall probe for scanning magnetic microscopy. *Sensors Actuators A* **129,** 212–215 (2006).

8. Kunets, V. P. *et al.* Highly sensitive micro-Hall devices based on Al0.12In0.88Sb/InSb heterostructures. *J. Appl. Phys.* **98,** 14506 (2005).

9. Hara, T., Mihara, M., Toyoda, N. & Zama, M. Highly Linear Gaas Hall Devices Fabricated By Ion Implantation. *IEEE Trans. Electron Devices* **ED-29,** 78–82 (1982).

10. Bando, M. *et al.* High sensitivity and multifunctional micro-Hall sensors fabricated using InAlSb/InAsSb/InAlSb heterostructures. *J. Appl. Phys.* **105,** 07E909 (2009).

11. Heremans, J. Solid state magnetic field sensors and applications. *J. Phys. D. Appl. Phys.* **26,** 1149–1168 (1993).

12. Xu, H. *et al.* Batch-fabricated high-performance graphene Hall elements. *Sci. Rep.* **3,** 1207

(2013).

13. Panchal, V. *et al.* Small epitaxial graphene devices for magnetosensing applications. *J. Appl. Phys.* **111,** 07E509 (2012).

14. Karpiak, B., Dankert, A. & Dash, S. P. Gate-tunable Hall sensors on large area CVD graphene protected by h-BN with 1D edge contacts. *J. Appl. Phys.* **122,** 54506 (2017).

15. Alexander-Webber, J. A. *et al.* Encapsulation of graphene transistors and vertical device integration by interface engineering with atomic layer deposited oxide. *2D Mater.* **4,** 11008 (2017).

16. Dean, C. R. *et al.* Boron nitride substrates for high-quality graphene electronics. *Nat. Nanotechnol.* **5,** 722–726 (2010).

17. Dauber, J. *et al.* Ultra-sensitive Hall sensors based on graphene encapsulated in hexagonal boron nitride. *Appl. Phys. Lett.* **106,** 193501 (2015).

18. Wang, Z. *et al.* Encapsulated graphene-based Hall sensors on foil with increased sensitivity. *Phys. Status Solidi B* **253,** 2316–2320 (2016).

19. Joo, M.-K. *et al.* Large-Scale Graphene on Hexagonal-BN Hall Elements: Prediction of Sensor Performance without Magnetic Field. *ACS Nano* **10,** 8803–8811 (2016).

20. Huang, L. *et al.* Ultra-sensitive graphene Hall elements. *Appl. Phys. Lett.* **104,** 183106 (2014).

21. Ciuk, T. *et al.* Low-noise epitaxial graphene on SiC Hall effect element for commercial applications. *Appl. Phys. Lett.* **108,** 223504 (2016).

22. Xu, H. *et al.* Flicker noise and magnetic resolution of graphene hall sensors at low frequency. *Appl. Phys. Lett.* **103,** 112405 (2013).

23. Wang, Z., Shaygan, M., Otto, M., Schall, D. & Neumaier, D. Flexible Hall sensors based on graphene. *Nanoscale* **8,** 7683–7687 (2016).

24. Wang, L. *et al.* One-dimensional electrical contact to a two-dimensional material. *Science* **342,** 614 (2013).

25. Graphenea Inc. Available at: http://www.graphenea.com.

26. Graphene Supermarket. Available at: https://graphene-supermarket.com.

27. Kamalakar, V. M., Groenveld, C., Dankert, A. & Dash, S. P. Long distance spin communication in chemical vapour deposited graphene. *Nat. Commun.* **6,** 6766 (2015).

28. Ferrari, A. ., Meyer, J. C., Scardaci, C., Casiraghi, C. & Lazzeri, M. Raman Spectrum of

Graphene and Graphene Layers. *Phys. Rev. Lett.* **97,** 187401 (2006).

29. Gorbachev, R. V. *et al.* Hunting for monolayer boron nitride: Optical and raman signatures. *Small* **7,** 465–468 (2011).

30. Kim, K. K. *et al.* Synthesis and Characterization of Hexagonal Boron Nitride Film as a Dielectric Layer for Graphene Devices. *ACS Nano* **6,** 8583–8590 (2012).

31. Jönsson-Åkerman, B. J. *et al.* Reliability of normal-state current–voltage characteristics as an indicator of tunnel-junction barrier quality. *Appl. Phys. Lett.* **77,** 1870 (2000).

32. Roy, S. S., Safron, N. S., Wu, M. & Arnold, M. S. Evolution, kinetics, energetics, and environmental factors of graphene degradation on silicon dioxide. *Nanoscale* **7,** 6093–6103 (2015).

33. Yang, Y., Brenner, K. & Murali, R. The influence of atmosphere on electrical transport in graphene. *Carbon N. Y.* **50,** 1727–1733 (2012).

34. Tang, S. *et al.* Nucleation and growth of single crystal graphene on hexagonal boron nitride. *Carbon N. Y.* **50,** 329–331 (2012).

35. Tang, S. *et al.* Silane-catalysed fast growth of large single-crystalline graphene on hexagonal boron nitride. *Nat. Commun.* **6,** 6499 (2015).

36. Tang, S. *et al.* Precisely aligned graphene grown on hexagonal boron nitride by catalyst free chemical vapor deposition. *Sci. Rep.* **3,** 2666 (2013).

# Supplementary materials

## Hall sensors batch-fabricated on all-CVD h-BN/graphene/h-BN heterostructures

André Dankert[†,*], Bogdan Karpiak[*], Saroj P. Dash[‡]

*Department of Microtechnology and Nanoscience, Chalmers University of Technology, SE-41296, Göteborg, Sweden.*

**S1. Characterization of all-CVD 2D materials and heterostructures.**

The 2D materials used in Hall sensor fabrication are grown by CVD method over large area. The CVD grown graphene samples were obtained from Graphenea and CVD h-BN from Graphene Supermarket. The Raman spectrum of graphene on $SiO_2$/Si substrate show the G and 2D peaks at 1597 cm$^{-1}$ and 2652 cm$^{-1}$ (Fig. S1a), with a small D peak[1]. At some areas bi-layer patches are observed. The grain size of the CVD graphene is mostly between 1-5 μm range.

Figure S1b shows Raman of multilayer h-BN with peak[2] at 1357 cm$^{-1}$ for selective places. The peak typical for h-BN is not present at most of the areas, pointing to structural disorders. As already known from TEM characterization of such multilayer h-BN by Kim et al[3], the films are polycrystalline in nature and the c-axis of the crystallites points to random directions. The EELS studies by Kim et al. also indicated that the h-BN film is having sp$^2$ bonds and 1:1 stoichiometry of B and N atoms.

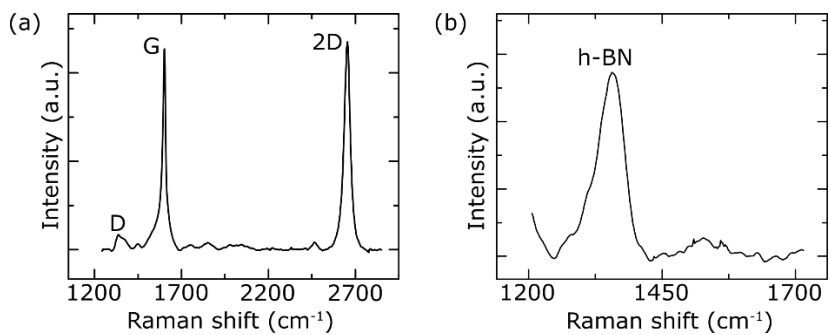

**Supplementary Figure S1.** *Raman spectra of (a) CVD graphene, (b) CVD h-BN.*

Corresponding authors: [†]andre.dankert@chalmers.se; [‡]saroj.dash@chalmers.se

* These authors contributed equally.

## S2. AFM of h-BN after annealing

To investigate the influence of annealing on the quality of CVD h-BN, the sample with h-BN after transfer on Si/SiO$_2$ substrate was put in a furnace at 400 °C in Ar/H$_2$ atmosphere for 8 hours. After the annealing, the topography of the surface was investigated by AFM (Fig. S2). The surface roughness of CVD h-BN was found to be ~2 nm, similar to the one before annealing (see Fig. 2b in the main text).

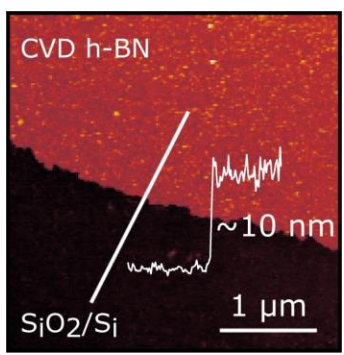

**Supplementary Figure S2.** *AFM image and thickness profile of CVD h-BN on SiO$_2$/Si wafer after annealing.*

## S3. Magnetic resolution of all-CVD h-BN/graphene/h-BN Hall sensors

From noise spectral measurements on fully-encapsulated graphene Hall sensor with Cr/Au contacts (Fig. S3) the minimum magnetic resolution ($B_{min}=S_V^{0.5}/(S_I I)$, where $S_V$ is the noise power spectral density) was found to be 0.4 mT/Hz$^{0.5}$ at frequency of 1 kHz, which is significantly lower in comparison to previous reports. Such difference could be attributed to the quality of h-BN/graphene/h-BN heterostructures after large-area wet transfer process from the CVD growth substrate.

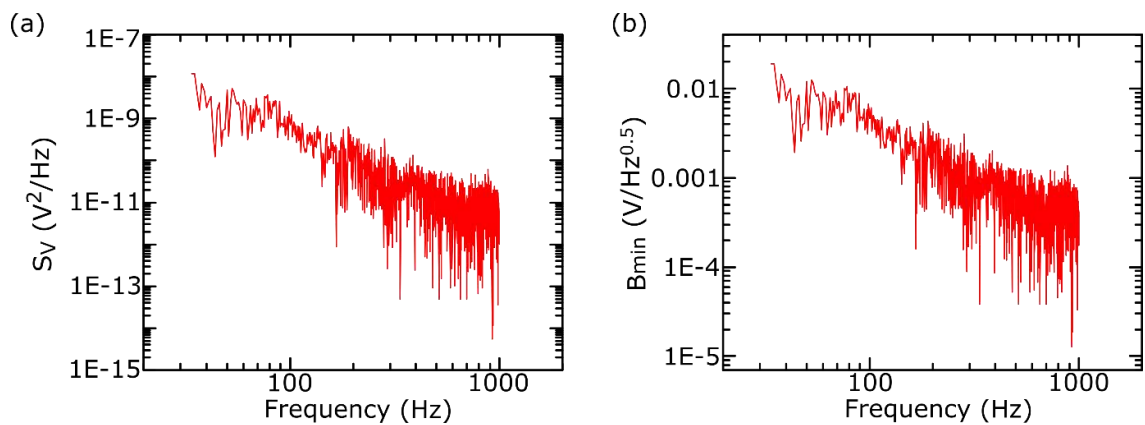

**Supplementary Figure S3.** *Noise characterization of all-CVD h-BN/graphene/h-BN Hall sensors. (a) Noise power spectral density $S_V$ and (b) minimum magnetic resolution $B_{min}$.*

## S4. Graphene Hall sensors prepared on $SiO_2$ substrate with and without exfoliated h-BN encapsulation

To compare the performance of Hall sensors fabricated on large area all-CVD heterostructures prepared by wet-transfer technique, we fabricated another control samples with CVD graphene on Si/$SiO_2$ substrate with and without top encapsulation by exfoliated h-BN (Fig. S4). From the Hall measurements (Figs. S3b and S3d) the Hall mobility was found to be around 150 cm$^2$/Vs$^{-1}$, hole doping concentration ~$10^{12}$ cm$^{-2}$. The current-related sensitivity $S_I$ in device with h-BN encapsulation (363 V/AT) was found to be 3 to 4 times higher than in all-CVD heterostructure-based Hall sensors, while unencapsulated devices showed $S_I$ in the range 100-200 V/AT after fabrication, degrading down to 50 V/AT after two months.

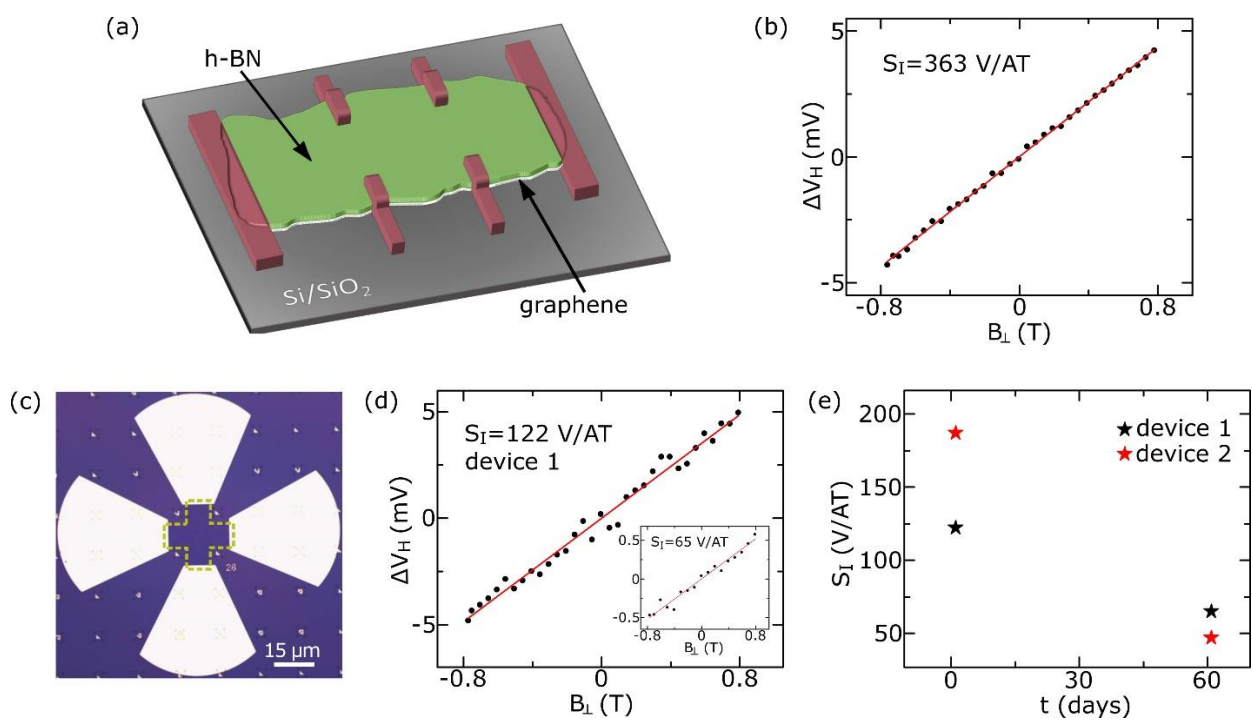

***Supplementary Figure S4.*** *CVD graphene Hall sensors fabricated on Si/$SiO_2$ substrate. (a) Schematic representation of the device with encapsulation by exfoliated h-BN flakes. (b) Hall voltage response of the device as a function of perpendicular magnetic field at room temperature (background is subtracted). (c) Optical microscope picture of the device without encapsulation. Graphene edges are marked by yellow dashed line. (d) Hall voltage response of the new unencapsulated device (main panel) and 2 months after fabrication (inset). Linear background is subtracted. (e) Time-dependent current-related sensitivity plot for the two unencapsulated CVD graphene samples.*

**References**

1. Ferrari, A. ., Meyer, J. C., Scardaci, C., Casiraghi, C. & Lazzeri, M. Raman Spectrum of Graphene and Graphene Layers. *Phys. Rev. Lett.* **97,** 187401 (2006).

2. Gorbachev, R. V. *et al.* Hunting for monolayer boron nitride: Optical and raman signatures. *Small* **7,** 465–468 (2011).

3. Kim, K. K. *et al.* Synthesis and Characterization of Hexagonal Boron Nitride Film as a Dielectric Layer for Graphene Devices. *ACS Nano* **6,** 8583–8590 (2012).